\documentclass[11pt]{article}
\usepackage{amssymb,amsmath,amsfonts}
\usepackage{graphicx}
\usepackage{graphics}
\usepackage{eepic,epsfig}
\usepackage{dcolumn}

\makeatother

\begin{document}

\title{Topological Casimir effect in models with helical compact dimensions}
\author{R. M. Avagyan$^{1,2}$, \thinspace A. A. Saharian$^{1,2}$\thanks{%
Corresponding author, E-mail: saharian@ysu.am}, D. H. Simonyan$^{1}$%
,\thinspace G. H. Harutyunyan$^{1}$ \\
\\
\textit{$^1$Institute of Physics, Yerevan State University,}\\
\textit{1 Alex Manoogian Street, 0025 Yerevan, Armenia} \vspace{0.3cm}\\
\textit{$^2$Institute of Applied Problems of Physics NAS RA,}\\
\textit{25 Hrachya Nersissyan Street, 0014 Yerevan, Armenia}}
\maketitle

\begin{abstract}
We investigate the influence of the helical compactification of spatial
dimension on the local properties of the vacuum state for a charged scalar
field with general curvature coupling parameter. A general background
geometry is considered with rotational symmetry in the subspace with the
coordinates appearing in the helical periodicity condition. It is shown that
by a coordinate transformation the problem is reduced to the problem with
standard quasiperiodicity condition in the same local geometry and with the
effective compactification radius determined by the length of the compact
dimension and the helicity parameter. As an application of the general
procedure we have considered locally de Sitter spacetime with a helical
compact dimension. By using the Hadamard function for the Bunch-Davies
vacuum state, the vacuum expectation values of the field squared, current
density, and energy-momentum tensor are studied. The topological
contributions are explicitly separated and their asymptotics are described
at early and late stages of cosmological expansion. An important difference,
compared to the problem with quasiperiodic conditions, is the appearance of
the nonzero off-diagonal component of the energy-momentum tensor and of the
component of the current density along the uncompact dimension.
\end{abstract}

\bigskip


\bigskip

\section{Introduction}

The topological effects play an important role in various fields of physics.
The latter include high-energy models with compact extra dimensions and
different types of condensed matter systems. As examples we mention here the
Kaluza-Klein type theories in supergravity and in string theories and
different types of topological structures of 2D materials. In field theories
formulated on background of spacetimes with nontrivial topology, in addition
to the field equations, periodicity conditions have to be imposed on the
fields along compact dimensions. As a consequence, the local physical
characteristics of fields depend on the global properties of the background
geometry. In particular, that is the case for the vacuum state of quantum
fields. In models with compact dimensions the influence of the periodicity
conditions on the properties of quantum vacuum is similar to that induced by
boundary conditions on the field operator in the Casimir effect and is known
as the topological Casimir effect (for reviews see \cite{Most97}-\cite%
{Khan14}). It has been investigated for different topological classes and
background geometries. The interest is motivated by applications in theories
with extra dimensions as a stabilization mechanism for moduli fields, in
cosmology as a possible source of dark energy driving the accelerated
expansion of the Universe, and in condensed matter physics as a source of
generation of ground state stresses and currents. Among other implications
of compact dimensions we can mention here new mechanisms for symmetry
breaking, the generation of topological mass in field theories and different
types of instabilities (see, for example, references \cite{Duff86}-\cite%
{Abre13}).

An interesting feature in theories with compact dimensions is the
possibility of inequivalent field configurations with different periodicity
conditions \cite{Most97,Isha78,Bana79}. The different conditions lead to
different physical consequences. Among the interesting directions in the
studies of the topological Casimir effect is the dependence of the physical
characteristics of the vacuum state on the periodicity conditions in the
compact subspace. The most popular conditions in the literature correspond
to periodic and antiperiodic fields (untwisted and twisted fields). They are
special cases of more general periodicity conditions for charged fields with
general phases. For the values of the phase different from 0 and $\pi $
vacuum currents appear along compact dimensions. Those currents have been
studied in \cite{Bell10}-\cite{Bell20AdS} for locally Minkowski, de Sitter
(dS) and anti-de Sitter (AdS) spacetimes (for a review see \cite{Saha24}).
More general helical conditions include an additional shift along uncompact
dimensions \cite{Dien01,Dien02}. The vacuum energy in models with helical
conditions along compact dimensions with zero value of the phase has been
studied in \cite{Feng10}-\cite{Fari22}. The current density in the case of
general phase is discussed in \cite{Saha23}.

In the present paper we show that the characteristics of the vacuum in
problems with helical periodicity conditions can be generated by using the
corresponding results for standard quasiperiodicity conditions by a
coordinate transformation depending on the length of compact dimension and
the helicity parameter. The organization of the paper is as follows. In the
next section the problem setup is presented. The coordinate transformation
is described and the connection between the vacuum expectation values (VEVs)
is given. As an example of general procedure, in Section \ref{sec:dS}, a
locally dS background geometry is considered. The expressions of the
Hadamard function, for the VEVs of the field squared, current density and
the energy-momentum tensor are presented. The main results of the paper are
summarized in Section \ref{sec:Conc}.

\section{Problem setup and coordinate transformation}

\label{sec:Setup}

Let us consider the background geometry described by the $(D+1)$-dimensional
line element $ds^{2}=g_{ik}dx^{i}dx^{k}$, where%
\begin{equation}
g_{ik}=g_{ik}(x_{\perp
}^{l}),\;g_{l,D-1}=g_{l,D}=0,\;g_{D-1,D-1}=g_{DD},\;x_{\perp
}^{l}=(x^{0},x^{1},\ldots ,x^{D-2}),  \label{gik}
\end{equation}%
with $l=0,1,\ldots ,D-2$. It will be assumed that the spatial coordinate $%
x^{D}$ is compactified to a circle with the length $a$, $0\leq x^{D}\leq a$
and for the coordinate $x^{D-1}$ one has $-\infty <x^{D-1}<+\infty $. No
specific conditions will be imposed on the geometry and topology of the
subspace covered by the coordinates $x_{\perp }^{l}$. We discuss the
dynamics of a scalar field $\varphi (x)$ with curvature coupling parameter $%
\xi $, governed by the equation%
\begin{equation}
\left( g^{ik}D_{i}D_{k}+\xi R+m^{2}\right) \varphi (x)=0,  \label{Feq}
\end{equation}%
where $D_{i}=\nabla _{i}+ieA_{i}$ is the gauge-covariant derivative and $e$
is the coupling between the scalar and gauge fields. Since the background
space has non-trivial topology, in addition to the field equation one should
specify the periodicity conditions along compact dimensions. In the subspace 
$(x^{D-1},x^{D})$ we impose helical periodicity condition

\begin{equation}
\varphi (x_{\perp }^{l},x^{D-1},x^{D}+a)=e^{i\alpha _{p}}\varphi (x_{\perp
}^{l},x^{D-1}+h,x^{D}),  \label{hc}
\end{equation}%
with the helicity parameter $h$ and constant phase $\alpha _{p}$. In the
special case $h=0$ the relation (\ref{hc}) reduces to a generic
quasiperiodicity condition.

Here we consider a simple configuration of the gauge field with $%
A_{D-1},A_{D}=\mathrm{const}$. These constant components of the gauge field
can be excluded from the field equation by the gauge transformation 
\begin{equation}
A_{i}=A_{i}^{\prime }+\partial _{i}\omega ,\text{ }\varphi (x)=e^{-ie\omega
}\varphi ^{\prime }(x),\text{ }\omega =A_{D-1}x^{D-1}+A_{D}x^{D}.
\label{gtr}
\end{equation}%
In the new gauge one has $A_{i}^{\prime }=0$ and $D_{i}^{\prime }=\nabla
_{i} $ for $l=D-1,D$. Now the condition (\ref{hc}) takes the form

\begin{equation}
\varphi ^{\prime }(x_{\perp }^{l},x^{D-1},x^{D}+a)=e^{i\tilde{\alpha}%
_{p}}\varphi ^{\prime }(x_{\perp }^{l},x^{D-1}+h,x^{D}),  \label{hcn}
\end{equation}%
with the new phase

\begin{equation}
\tilde{\alpha}_{p}=\alpha _{p}-eA_{D-1}h+eA_{D}a.  \label{alft}
\end{equation}%
The physical characteristics will depend on the quantities $\alpha _{p}$, $%
A_{D-1}$, $A_{D}$ in the form of the combination.

In quantum field theory the periodicity conditions imposed on the field
operator modify the spectrum of vacuum fluctuations and the vacuum
expectation values of physical observables are shifted by amount that
depends on the parameters of the compactification (the topological Casimir
effect \cite{Most97}-\cite{Khan14}). These effects for the quasiperiodicity
conditions, corresponding to the zero value of the helicity parameter, $h=0$%
, have been widely investigated in the literature for different local
geometries. Simple geometries with helical conditions in the case of zero
phase were discussed in \cite{Feng10}-\cite{Fari22}. In the discussion below
we will show that the results for the helical conditions can be obtained by
an appropriate coordinate transformations from the formulas for
quasiperiodicity conditions.

The helical condition identifies the spacetime points with the coordinates $%
P_{(0,a)}=(x_{\perp }^{l},x^{D-1},x^{D}+a)$ and $P_{(h,0)}=(x_{\perp
}^{l},x^{D-1}+h,x^{D})$. Let us introduce new coordinates $\bar{x}^{i}$ in
accordance with $\bar{x}^{l}=x^{l}$ for $l=0,1,\ldots ,D-2$, and%
\begin{eqnarray}
\bar{x}^{D-1} &=&\frac{a}{\bar{a}}x^{D-1}+\frac{h}{\bar{a}}x^{D}+\frac{a}{%
\bar{a}}h.  \notag \\
\bar{x}^{D} &=&-\frac{h}{\bar{a}}x^{D-1}+\frac{a}{\bar{a}}x^{D}-\frac{h^{2}}{%
\bar{a}},  \label{xbar}
\end{eqnarray}%
where 
\begin{equation}
\bar{a}=\sqrt{a^{2}+h^{2}}.  \label{abar}
\end{equation}%
The inverse transformation reads%
\begin{eqnarray}
x^{D-1} &=&\frac{a}{\bar{a}}\bar{x}^{D-1}-\frac{h}{\bar{a}}\bar{x}^{D}-h. 
\notag \\
x^{D} &=&\frac{h}{\bar{a}}\bar{x}^{D-1}+\frac{a}{\bar{a}}\bar{x}^{D}.
\label{xinv}
\end{eqnarray}%
For the identification points in the coordinates $\bar{x}^{i}$ one has%
\begin{equation}
P_{(0,a)}=(\bar{x}_{\perp }^{l},\bar{x}^{D-1},\bar{x}^{D}+\bar{a}%
),\;P_{(h,0)}=(\bar{x}_{\perp }^{l},\bar{x}^{D-1},\bar{x}^{D}).  \label{Pbar}
\end{equation}%
The coordinate transformation (\ref{xbar}) is a combination of the rotation
by angle $\theta =\arctan (h/a)$, $0\leq \theta \leq \pi /2$, and the shift
of the origin to the point $x^{i}=(x_{\perp }^{l},-h,0)$. The metric tensor
is form-invariant under the transformation (\ref{xbar}).

Now we can reformulate the problem of the investigation of the VEVs for the
field $\varphi (x)$ with helical condition (\ref{hc}) in the coordinate
system $\bar{x}^{i}$. For the corresponding metric tensor we still have%
\begin{equation}
\bar{g}_{ik}=\bar{g}_{ik}(\bar{x}_{\perp }^{l}),\;\bar{g}_{l,D-1}=\bar{g}%
_{l,D}=0,\;\bar{g}_{D-1,D-1}=\bar{g}_{DD},  \label{gbar}
\end{equation}%
with $\bar{x}_{\perp }^{l}=x_{\perp }^{l}$. The field equation has the form (%
\ref{Feq}) with the replacements $g^{ik}\rightarrow \bar{g}^{ik}$ for the
metric tensor and $D_{i}\rightarrow \bar{D}_{i}=\bar{\nabla}_{i}+ie\bar{A}%
_{i}$ for the covariant derivative, where $\bar{A}_{i}=A_{i}$ for $%
i=0,1,\ldots ,D-2$, and 
\begin{eqnarray}
\bar{A}_{D-1} &=&\frac{a}{\bar{a}}A_{D-1}+\frac{h}{\bar{a}}A_{D},  \notag \\
\bar{A}_{D} &=&-\frac{h}{\bar{a}}A_{D-1}+\frac{a}{\bar{a}}A_{D}.
\label{Abar}
\end{eqnarray}%
In the new coordinates the periodicity condition takes the form%
\begin{equation}
\bar{\varphi}(\bar{x}_{\perp }^{l},\bar{x}^{D-1},\bar{x}^{D}+\bar{a}%
)=e^{i\alpha _{p}}\bar{\varphi}(\bar{x}_{\perp }^{l},\bar{x}^{D-1},\bar{x}%
^{D}),  \label{qpc}
\end{equation}%
which is a standard quasiperiodicity condition. This shows that we can use
the results for the VEVs in problems with quasiperiodicity condition (\ref%
{qpc}) in order to find the expectation values in problems with helical
conditions. Let us specify this procedure for the current density and the
energy-momentum tensor. The renormalized VEVs in the coordinate system $\bar{%
x}^{i}$ we denote by $\left\langle \bar{j}^{i}\right\rangle =\left\langle 
\bar{j}^{i}(\alpha _{p},\bar{A}_{l})\right\rangle $ and $\left\langle \bar{T}%
^{ik}\right\rangle =\left\langle \bar{T}^{ik}(\alpha _{p},\bar{A}%
_{l})\right\rangle $ for the current density and the energy-momentum tensor,
respectively. The corresponding expectation values $\left\langle
j^{i}\right\rangle =\left\langle j^{i}(\alpha _{p},A_{l})\right\rangle $ and 
$\left\langle T^{ik}\right\rangle =\left\langle T^{ik}(\alpha
_{p},A_{l})\right\rangle $ in the original problem with helical periodicity
condition (\ref{hc}) are obtained by the coordinate transformation $\bar{x}%
^{i}\rightarrow x^{i}$.

We start with the current density. Note that in the coordinate system $\bar{x%
}^{i}$ we can make a gauge transformation $\bar{A}_{i}=\bar{A}_{i}^{\prime }+%
\bar{\partial}_{i}\bar{\omega}$, $\bar{\varphi}(\bar{x})=e^{-ie\bar{\omega}}%
\bar{\varphi}^{\prime }(\bar{x})$ with the function $\bar{\omega}=\bar{A}%
_{D-1}\bar{x}^{D-1}$. In the new gauge one gets $\bar{A}_{D-1}^{\prime }=0$.
Both the field equation and the periodicity condition (\ref{qpc}) are
invariant under this gauge transformation and the physical results do not
depend on $\bar{A}_{D-1}$. In the gauge $\bar{A}_{D-1}=0$ the metric tensor,
the field equation and the periodicity condition in the coordinate system $%
\bar{x}^{i}$ are symmetric under the reflection $\bar{x}^{D-1}\rightarrow -%
\bar{x}^{D-1}$. Assuming that the vacuum state is also symmetric under this
reflection, we conclude that the component of the current density along the
coordinate direction $\bar{x}^{D-1}$ vanishes by the symmetry, $\left\langle 
\bar{j}^{D-1}\right\rangle =0$. In this case the components of the current
density in the coordinates $x^{i}$ are expressed as%
\begin{eqnarray}
\left\langle j^{i}\right\rangle &=&\left\langle \bar{j}^{i}\right\rangle
,\;i=0,1,\ldots ,D-2,  \notag \\
\left\langle j^{D-1}\right\rangle &=&-\frac{h\,\left\langle \bar{j}%
^{D}\right\rangle }{\sqrt{a^{2}+h^{2}}},\;\left\langle j^{D}\right\rangle =%
\frac{a\,\left\langle \bar{j}^{D}\right\rangle }{\sqrt{a^{2}+h^{2}}}.
\label{jD}
\end{eqnarray}%
and the vacuum current density has a nonzero component along the uncompact
dimension $x^{D-1}$ as well. The components along compact and uncompact
dimensions related by the helical condition are connected by the formula 
\begin{equation}
\left\langle j^{D-1}\right\rangle =-\frac{h}{a}\left\langle
j^{D}\right\rangle .  \label{jrel}
\end{equation}
This relation for the locally Minkowski bulk was obtained in \cite{Saha23}
by direct evaluation of the VEV using the corresponding mode functions.

Another important characteristic of the vacuum state is the expectation
value of the energy-momentum tensor. Again, assuming that the vacuum is
symmetric with respect to the reflection $\bar{x}^{D-1}\rightarrow -\bar{x}%
^{D-1}$, we conclude that $\left\langle \bar{T}^{i,D-1}\right\rangle =0$ for 
$i\neq D-1$. By using the transformation rule for the second rank tensor,
for the components of the energy-momentum tensor we get ($i,k=0,1,\ldots
,D-2 $)%
\begin{eqnarray}
\left\langle T^{ik}\right\rangle &=&\left\langle \bar{T}^{ik}\right\rangle
,\;i,k=0,1,\ldots ,D-2,  \notag \\
\left\langle T^{iD}\right\rangle &=&-\frac{a}{h}\left\langle
T^{i,D-1}\right\rangle =\frac{a\,\left\langle \bar{T}^{iD}\right\rangle }{%
\sqrt{a^{2}+h^{2}}},  \label{TiD}
\end{eqnarray}%
for the components with one or two indices in the subspace $%
(x^{0},x^{1},\ldots ,x^{D-2})$ and%
\begin{eqnarray}
\left\langle T^{D-1,D-1}\right\rangle &=&\frac{a^{2}\left\langle \bar{T}%
^{D-1,D-1}\right\rangle }{a^{2}+h^{2}}+\frac{h^{2}\left\langle \bar{T}%
^{DD}\right\rangle }{a^{2}+h^{2}},  \notag \\
\left\langle T^{D-1,D}\right\rangle &=&\frac{ah}{a^{2}+h^{2}}\left[
\left\langle \bar{T}^{D-1,D-1}\right\rangle -\left\langle \bar{T}%
^{DD}\right\rangle \right] ,  \notag \\
\left\langle T^{DD}\right\rangle &=&\frac{h^{2}\left\langle \bar{T}%
^{D-1,D-1}\right\rangle }{a^{2}+h^{2}}+\frac{a^{2}\left\langle \bar{T}%
^{DD}\right\rangle }{a^{2}+h^{2}},  \label{TDD}
\end{eqnarray}%
for the components in the subspace $(x^{D-1},x^{D})$.

Note that the condition (\ref{hc}) can also be interpreted as a helical
periodicity condition along the compact dimension $x^{D-1}$ with the length $%
h$, with the helicity parameter $a$ along the uncompact direction $x^{D}$,
and with the phase $-\alpha _{p}$. This shows that there is a duality
between the models with the sets $(a,h,\alpha _{p})$ and $(h,a,-\alpha _{p})$%
. In the dual models the roles of the dimensions $x^{D-1}$ and $x^{D}$ are
interchanged. The duality is also seen in the VEVs (\ref{jD}), (\ref{TiD}),
and (\ref{TDD}).

\section{Models with locally dS spacetime}

\label{sec:dS}

As an application of the general procedure described above let us consider a
background spacetime with local dS geometry. The dS spacetime is among the
most popular geometries in quantum field theory in curved spacetime. In
particular, that is motivated by important applications in inflationary
models of the early Universe and in models of accelerating expansion at
recent epoch. In inflationary coordinates the corresponding line element
reads%
\begin{equation}
ds^{2}=dt^{2}-e^{2t/\alpha }\sum_{i=1}^{D}(dx^{i})^{2},  \label{ds2}
\end{equation}%
where the constant $\alpha $ determines the curvature radius of the
spacetime. It is expressed in terms of the cosmological constant $\Lambda $
by the formula $\alpha ^{2}=D(D-1)/(2\Lambda )$. For the remaining spatial
dimensions we take $-\infty <x^{i}<+\infty $, $i=1,\ldots ,D-1$. Introducing
a conformal time $\tau $ in accordance with $\tau =-\alpha e^{-t/\alpha }$,
the line element is written in a conformally flat form%
\begin{equation}
ds^{2}=g_{ik}dx^{i}dx^{k}=\frac{\alpha ^{2}}{\eta ^{2}}\left[ d\tau
^{2}-\sum_{i=1}^{D}(dx^{i})^{2}\right] ,  \label{ds2b}
\end{equation}%
where $\eta =|\tau |$. For the scalar curvature in the field equation (\ref%
{Feq}) one has $R=D(D+1)/\alpha ^{2}$. The VEVs of the field squared and
energy-momentum tensor in the model with a single compact dimension and
periodic condition along it were studied in \cite{Saha08b}. The general case
of spatial topology $R^{p}\times (S^{1})^{q}$, $p+q=D$, has been discussed
in \cite{Bell08,Saha09}. The vacuum currents for quasiperiodic conditions
with general phases are investigated in \cite{Bell13}. For simplicity here
we consider the special case $p=D-1$ and $q=1$, assuming that the only
compact dimension corresponds to the coordinate $x^{D}$ along which the
quantum scalar field obeys the condition (\ref{hc}). In the discussion below
we will work in the coordinate system (\ref{ds2b}).

\subsection{Hadamard function and the VEVs of the field squared and current
density}

The local characteristics of the vacuum state $\left\vert 0\right\rangle $
for a quantum scalar field $\varphi (x)$ are obtained from the two-point
functions. They describe the correlations of zero-point fluctuations at
different spacetime points $x$ and $x^{\prime }$. As a two-point function we
will take the Hadamard function defined as the VEV%
\begin{equation}
G(x,x^{\prime })=\left\langle 0\right\vert \varphi (x)\varphi ^{\dagger
}(x^{\prime })+\varphi ^{\dagger }(x^{\prime })\varphi (x)\left\vert
0\right\rangle .  \label{G1}
\end{equation}%
For dS spacetime different vacuum states have been considered in the
literature. Among them the Bunch-Davies vacuum is distinguished by the
following two properties: it is maximally symmetric and is reduced to the
Minkowski vacuum in flat spacetime in the slow expansion limit. Here we
assume that the field $\varphi (x)$ is prepared in the Bunch-Davies vacuum
state. The Hadamard function $\bar{G}(\bar{x},\bar{x}^{\prime })$ in the
problem at hand for the coordinates $\bar{x}^{i}$ is obtained from the
expression in \cite{Bell13} as a special case. Transforming to the
coordinates $x^{i}$ we get 
\begin{eqnarray}
G(x,x^{\prime }) &=&\frac{4\left( \eta \eta ^{\prime }\right) ^{D/2}}{\left(
2\pi \right) ^{D/2+1}\alpha ^{D-1}}\int_{0}^{\infty }dz\,z\left[ I_{-\nu
}(\eta z)K_{\nu }(\eta ^{\prime }z)+K_{\nu }(\eta z)I_{\nu }(\eta ^{\prime
}z)\right]   \notag \\
&&\times \sum_{n=-\infty }^{\infty }e^{-in\tilde{\alpha}_{p}}\frac{%
f_{D/2-1}(z\sqrt{|\Delta \mathbf{x}_{D}|^{2}+n^{2}(a^{2}+h^{2})+2n\left(
a\Delta x^{D}-h\Delta x^{D-1}\right) })}{[|\Delta \mathbf{x}%
_{D}|^{2}+n^{2}(a^{2}+h^{2})+2n\left( a\Delta x^{D}-h\Delta x^{D-1}\right)
]^{D/2-1}}.  \label{Gb}
\end{eqnarray}%
where $\mathbf{x}_{D}=(x^{1},\ldots ,x^{D})$, $\Delta \mathbf{x}_{D}=\mathbf{%
x}_{D}-\mathbf{x}_{D}^{\prime }$, $\Delta x^{i}=x^{i}-x^{\prime i}$, $I_{\nu
}(z)$ and $K_{\nu }(z)$ are the modified Bessel functions \cite{Abra72} with
the order 
\begin{equation}
\nu =\left[ D^{2}/4-D(D+1)\xi -m^{2}\alpha ^{2}\right] ^{1/2}.  \label{nu}
\end{equation}%
The function $f_{\mu }(y)$ in (\ref{Gb}) is defined by $f_{\mu }(y)=y^{\mu
}K_{\mu }(y)$. The $n=0$ term in (\ref{Gb}) corresponds to the Hadamard
function $G_{\mathrm{dS}}(x,x^{\prime })$ in the dS spacetime without
compactification and the remaining part is induced by the helical
compactification. The expression for $G_{\mathrm{dS}}(x,x^{\prime })$ in
terms of the hypergeometric function is well known from the literature.

Given the Hadamard function, the VEVs of physical observables are obtained
taking the coincidence limit of the arguments of the Hadamard function or
its derivatives. We start with the VEV of the field squared $\left\langle
\varphi \varphi ^{\dagger }\right\rangle =\left\langle 0\right\vert \varphi
\varphi ^{\dagger }\left\vert 0\right\rangle $. It is obtained in the limit $%
\left\langle \varphi \varphi ^{\dagger }\right\rangle =\lim_{x^{\prime
}\rightarrow x}G(x,x^{\prime })/2$. This limit is divergent and a
renormalization is required. The compactification scheme under consideration
does not change the local geometry and the divergences are the same as in
the dS spacetime without compactification. The corresponding part in the
Hadamard function (\ref{Gb}) is presented by the $n=0$ term. Separating the
topological contribution and taking the coincidence limit the VEV is
decomposed as 
\begin{equation}
\langle \varphi \varphi ^{\dagger }\rangle =\langle \varphi \varphi
^{\dagger }\rangle _{\mathrm{dS}}+\langle \varphi \varphi ^{\dagger }\rangle
_{\mathrm{c}},  \label{phi2d}
\end{equation}%
where the renormalized VEV $\langle \varphi \varphi ^{\dagger }\rangle _{%
\mathrm{dS}}$ in dS spacetime has been already studied in the literature. By
the maximal symmetry of the Bunch-Davies vacuum state, it does not depend on
spacetime coordinates. The topological contribution $\langle \varphi \varphi
^{\dagger }\rangle _{\mathrm{c}}$ is directly obtained from the part in (\ref%
{Gb}) with $n\neq 0$ in the coincidence limit:%
\begin{equation}
\langle \varphi \varphi ^{\dagger }\rangle _{\mathrm{c}}=\frac{4\alpha
^{1-D}\eta ^{D-2}}{(2\pi )^{\frac{D}{2}+1}(a^{2}+h^{2})^{\frac{D}{2}-1}}%
\int_{0}^{\infty }dz\,zF_{\nu }(z)\sum_{n=1}^{\infty }\frac{\cos \left( n%
\tilde{\alpha}_{p}\right) }{n^{D-2}}f_{\frac{D}{2}-1}(y_{n}),  \label{phi2}
\end{equation}%
with the notations%
\begin{equation}
y_{n}=\frac{nz}{\eta }\sqrt{a^{2}+h^{2}},  \label{yn}
\end{equation}%
and%
\begin{equation}
F_{\nu }(z)=\left[ I_{-\nu }(z)+I_{\nu }(z)\right] K_{\nu }(z).  \label{Fnu}
\end{equation}%
The VEV (\ref{phi2}) is an even function of the phase $\tilde{\alpha}_{p}$.
This corresponds to the periodicity with respect to the magnetic flux
enclosed by the compact dimension, with the period equal to the flux
quantum. In addition, the mean field squared $\langle \varphi ^{2}\rangle _{%
\mathrm{c}}$ is invariant under the change $(a,h,\alpha _{p})\rightarrow
(h,a,-\alpha _{p})$. This is a manifestation of the duality mentioned above.

For a charged scalar field the operator of the current density is given by%
\begin{equation}
j_{l}=ie[\varphi ^{\dagger }D_{l}\varphi -(D_{l}\varphi )^{\dagger }\varphi
].  \label{jmu}
\end{equation}%
The corresponding VEV can be obtained in two different ways. The first one
corresponds to the limiting transition (in the gauge where $A_{i}=0$)%
\begin{equation}
\left\langle j_{l}\right\rangle =\frac{i}{2}e\lim_{x^{\prime }\rightarrow
x}(\partial _{l}-\partial _{l}^{\prime })G(x,x^{\prime }),  \label{jmuG}
\end{equation}%
by using the Hadamard function (\ref{Gb}). Note that the limit in the
right-hand side of (\ref{jmuG}) with the dS Hadamard function vanishes and
the renormalization for the current density is not required. In the second
way, the vacuum current density is obtained from the corresponding result
for quasiperiodic condition, given in \cite{Bell13}, by the coordinate
transformation (\ref{jD}). For the nonzero components we get 
\begin{equation}
\left\langle j^{D}\right\rangle =\frac{8e\alpha ^{-D-1}a\eta ^{D}}{\left(
2\pi \right) ^{\frac{D}{2}+1}(a^{2}+h^{2})^{\frac{D}{2}}}\int_{0}^{\infty
}dz\,\,zF_{\nu }(z)\sum_{n=1}^{\infty }\frac{\sin (n\tilde{\alpha}_{p})}{%
n^{D-1}}f_{\frac{D}{2}}(y_{n}),  \label{jDdS}
\end{equation}%
and $\left\langle j^{D-1}\right\rangle =-h\left\langle j^{D}\right\rangle /a$%
. Here, $y_{n}$ is defined by (\ref{yn}). The physical components of the
current density, denoted here by $\left\langle j_{\mathrm{p}%
}^{l}\right\rangle $, are connected to the contravariant components by the
relation $\left\langle j_{\mathrm{(p)}}^{l}\right\rangle =\left( \alpha
/\eta \right) \left\langle j^{l}\right\rangle $. The components $%
\left\langle j^{D}\right\rangle $ and $\left\langle j^{D-1}\right\rangle $
are odd functions of the phase $\tilde{\alpha}_{p}$. In particular, the
current density vanishes for half-integer values of the parameter $\tilde{%
\alpha}_{p}$. In agreement with the duality mentioned at the end of the
previous section, the current densities are invariant under the change $%
(a,h,\alpha _{p})\rightarrow (h,a,-\alpha _{p})$ with the change of the
roles of the coordinates $(x^{D-1},x^{D})\rightarrow (x^{D},x^{D-1})$.

\subsection{Vacuum energy-momentum tensor}

Finally, we turn to the VEV of the energy-momentum tensor. In the gauge with 
$A_{i}=0$ it is obtained from the Hadamard function (\ref{Gb}) with the help
of the formula (again, in the gauge with zero gauge potential)%
\begin{equation}
\langle T_{ik}\rangle =\lim_{x^{\prime }\rightarrow x}\partial _{i}\partial
_{k}^{\prime }G(x,x^{\prime })+\left[ \left( \xi -\frac{1}{4}\right)
g_{ik}\nabla _{l}\nabla ^{l}-\xi \nabla _{i}\nabla _{k}-\xi R_{ik}\right]
\langle \varphi ^{2}\rangle ,  \label{Tikg}
\end{equation}%
where $R_{ik}=Dg_{ik}/\alpha ^{2}$ is the Ricci tensor for the dS spacetime.
Alternatively, the VEV is derived by the coordinate transformation from the
results in the coordinate system $\bar{x}^{i}$ with standard periodicity
condition. The corresponding formulas in the special case $\tilde{\alpha}%
_{p}=0$ are obtained from the results of \cite{Bell08}. Generalizing for $%
\tilde{\alpha}_{p}\neq 0$ and using the transformation rules (\ref{TiD}) and
(\ref{TDD}) one finds 
\begin{equation}
\langle T_{i}^{k}\rangle =\langle T_{i}^{k}\rangle _{\mathrm{dS}}+\langle
T_{i}^{k}\rangle _{\mathrm{c}},  \label{TikComp}
\end{equation}%
where $\langle T_{i}^{k}\rangle _{\mathrm{dS}}=\mathrm{const}\cdot \delta
_{i}^{k}$ is the corresponding VEV in the dS spacetime without
compactification. The topological contribution for the vacuum energy density
reads%
\begin{equation}
\langle T_{0}^{0}\rangle _{\mathrm{c}}=\frac{2\alpha ^{-1-D}\eta ^{D-2}}{%
(2\pi )^{\frac{D}{2}+1}(a^{2}+h^{2})^{\frac{D}{2}-1}}\sum_{n=1}^{\infty }%
\frac{\cos \left( n\tilde{\alpha}_{p}\right) }{n^{D-2}}\int_{0}^{\infty
}dz\,zF^{(0)}(z)f_{\frac{D}{2}-1}(y_{n}),  \label{T00}
\end{equation}%
with the notation%
\begin{equation}
F^{(0)}(z)=\frac{1}{2}z\left[ zF_{\nu }^{\prime }(z)\right] ^{\prime
}+D\left( \frac{1}{2}-2\xi \right) zF_{\nu }^{\prime }(z)+2\left(
m^{2}\alpha ^{2}-z^{2}\right) F_{\nu }(z),  \label{F0}
\end{equation}%
where the prime stands for the derivative with respect to $z$. For the
vacuum stresses along the directions $x^{i}$, with $i=1,2,\ldots ,D-2$, one
gets (no summation over $i$)%
\begin{equation}
\langle T_{i}^{i}\rangle _{\mathrm{c}}=\frac{2\alpha ^{-1-D}\eta ^{D-2}}{%
(2\pi )^{\frac{D}{2}+1}(a^{2}+h^{2})^{\frac{D}{2}-1}}\sum_{n=1}^{\infty }%
\frac{\cos \left( n\tilde{\alpha}_{p}\right) }{n^{D-2}}\int_{0}^{\infty
}dz\,z\left[ F(z)f_{\frac{D}{2}-1}(y_{n})-\frac{2\eta ^{2}F_{\nu }(z)}{%
n^{2}(a^{2}+h^{2})}f_{\frac{D}{2}}(y_{n})\right] ,  \label{Tii}
\end{equation}%
where%
\begin{equation}
F(z)=\left( 2\xi -\frac{1}{2}\right) z\left[ zF_{\nu }^{\prime }(z)\right]
^{\prime }+\left[ 2(D+1)\xi -\frac{D}{2}\right] zF_{\nu }^{\prime }(z).
\label{Fz}
\end{equation}

Now we turn to the components of the topological part in the vacuum
energy-momentum tensor with one or two indices in the subspace $%
(x^{D-1},x^{D})$. The off-diagonal components $\langle T_{D}^{i}\rangle _{%
\mathrm{c}}$, with $i=1,2,\ldots ,D-2$, vanish: $\langle T_{D}^{i}\rangle _{%
\mathrm{c}}=0$. For the diagonal components in the subspace $(x^{D-1},x^{D})$
we find%
\begin{eqnarray}
\left\langle T_{D-1}^{D-1}\right\rangle _{\mathrm{c}} &=&\frac{2\alpha
^{-1-D}\eta ^{D-2}}{(2\pi )^{\frac{D}{2}+1}(a^{2}+h^{2})^{\frac{D}{2}-1}}%
\sum_{n=1}^{\infty }\frac{\cos \left( n\tilde{\alpha}_{p}\right) }{n^{D-2}}%
\int_{0}^{\infty }dz\,z\left\{ F(z)f_{\frac{D}{2}-1}(y_{n})\right.  \notag \\
&&\left. -\frac{2\eta ^{2}F_{\nu }(z)}{n^{2}(a^{2}+h^{2})}\left[ \left( 1-%
\frac{Dh^{2}}{a^{2}+h^{2}}\right) f_{\frac{D}{2}}(y_{n})-h^{2}\frac{%
y_{n}^{2}f_{\frac{D}{2}-1}(y_{n})}{a^{2}+h^{2}}\right] \right\} ,
\label{TDm}
\end{eqnarray}%
and 
\begin{eqnarray}
\left\langle T_{D}^{D}\right\rangle _{\mathrm{c}} &=&\frac{2\alpha
^{-1-D}\eta ^{D-2}}{(2\pi )^{\frac{D}{2}+1}(a^{2}+h^{2})^{\frac{D}{2}-1}}%
\sum_{n=1}^{\infty }\frac{\cos \left( n\tilde{\alpha}_{p}\right) }{n^{D-2}}%
\int_{0}^{\infty }dz\,z\left\{ F(z)f_{\frac{D}{2}-1}(y_{n})\right.  \notag \\
&&\left. -\frac{2\eta ^{2}F_{\nu }(z)}{n^{2}(a^{2}+h^{2})}\left[ \left( 1-%
\frac{Da^{2}}{a^{2}+h^{2}}\right) f_{\frac{D}{2}}(y_{n})-a^{2}y_{n}^{2}\frac{%
f_{\frac{D}{2}-1}(y_{n})}{a^{2}+h^{2}}\right] \right\} ,  \label{TD}
\end{eqnarray}%
with $y_{n}$ from (\ref{yn}). In addition to the diagonal components, the
helical periodicity condition induces a nonzero off-diagonal component $%
\left\langle T_{D}^{D-1}\right\rangle _{\mathrm{c}}$. It is obtained from
the diagonal components in the coordinate system $\bar{x}^{i}$ by the
transformation given by (\ref{TDD}): 
\begin{equation}
\left\langle T_{D}^{D-1}\right\rangle _{\mathrm{c}}=\frac{-4\alpha
^{-1-D}\eta ^{D}ah}{(2\pi )^{\frac{D}{2}+1}(a^{2}+h^{2})^{\frac{D}{2}+1}}%
\sum_{n=1}^{\infty }\frac{\cos \left( n\tilde{\alpha}_{p}\right) }{n^{D}}%
\int_{0}^{\infty }dz\,zF_{\nu }(z)\left[ y_{n}^{2}f_{\frac{D}{2}%
-1}(y_{n})+Df_{\frac{D}{2}}(y_{n})\right] .  \label{Toff}
\end{equation}%
All the components of the vacuum energy-momentum tensor are even functions
of $\tilde{\alpha}_{p}$. Note that the parameter $\nu $ defined by (\ref{nu}%
) can be either nonnegative real number or purely imaginary. The integral
representations given above are valid in the range $\mathrm{Re}\,\nu <1$.
This restriction follows from the condition of the convergence of the
integrals over $z$ in the lower limit.

It can be explicitly checked that the topological part of the vacuum
energy-momentum tensor obeys the trace relation%
\begin{equation}
\langle T_{i}^{i}\rangle _{\mathrm{c}}=\left[ D(\xi -\xi _{D})\nabla
_{l}\nabla ^{l}+m^{2}\right] \langle \varphi ^{2}\rangle _{\mathrm{c}},
\label{Trace}
\end{equation}%
where $\xi _{D}=(D-1)/4D$ is the value of the curvature coupling parameter
for a conformally coupled scalar field. For a conformally coupled massless
field the topological contribution $\langle T_{i}^{k}\rangle _{\mathrm{c}}$
is traceless. The anomaly in the trace is contained in the pure dS part $%
\langle T_{i}^{k}\rangle _{\mathrm{dS}}$.

Note that the parameters $a$ and $h$ are the coordinate lengths. The
corresponding physical (proper) lengths measured by an observer at rest in
the coordinates $x^{i}$ are given by $a_{\mathrm{(p)}}=\alpha a/\eta $ and $%
h_{\mathrm{(p)}}=\alpha h/\eta $. The VEVs $\langle \varphi \varphi
^{\dagger }\rangle _{\mathrm{c}}$, $\left\langle j_{\mathrm{(p)}%
}^{l}\right\rangle $, and $\langle T_{i}^{k}\rangle _{\mathrm{c}}$ depend on 
$a$, $h$, and $\eta $ in the form of the ratios $a/\eta $ and $h/\eta $. The
latter are the proper lengths measured in units of the curvature radius $%
\alpha $.

\subsection{Conformally coupled massless field and the asymptotics}

For a conformally coupled massless field one has $\nu =1/2$ and $F_{\nu
}(z)=1/z$. The integrals are evaluated by using the formulae \cite{Prud2}%
\begin{equation}
\int_{0}^{\infty }dy\,f_{\frac{D}{2}}(y)=\int_{0}^{\infty }dy\,y^{2}f_{\frac{%
D}{2}-1}(y)=(D-1)\int_{0}^{\infty }dy\,f_{\frac{D}{2}-1}(y)=2^{\frac{D}{2}-1}%
\sqrt{\pi }\Gamma \left( \frac{D+1}{2}\right) .  \label{Ints}
\end{equation}%
For the VEVs of the field squared and physical component of the current
density one gets%
\begin{eqnarray}
\langle \varphi \varphi ^{\dagger }\rangle _{\mathrm{c}} &=&\frac{\Gamma
((D-1)/2)(\eta /\alpha )^{D-1}}{2\pi ^{\frac{D+1}{2}}(a^{2}+h^{2})^{\frac{D-1%
}{2}}}\sum_{n=1}^{\infty }\frac{\cos \left( n\tilde{\alpha}_{p}\right) }{%
n^{D-1}},  \notag \\
\left\langle j_{\mathrm{(p)}}^{D}\right\rangle _{\mathrm{c}} &=&\frac{%
2e\Gamma ((D+1)/2)(\eta /\alpha )^{D}a}{\pi ^{\frac{D+1}{2}}(a^{2}+h^{2})^{%
\frac{D+1}{2}}}\,\sum_{n=1}^{\infty }\frac{\sin (n\tilde{\alpha}_{p})}{n^{D}}%
,\;\left\langle j_{\mathrm{(p)}}^{D-1}\right\rangle _{\mathrm{c}}=-\frac{h}{a%
}\left\langle j_{\mathrm{(p)}}^{D}\right\rangle _{\mathrm{c}}.  \label{jDcc}
\end{eqnarray}%
The expressions for the energy density and stresses along the directions $%
x^{i}$, $i=1,2,\ldots ,D-2$, are simplified to (no summation over $%
i=0,1,\ldots ,D-2$)%
\begin{equation}
\langle T_{i}^{i}\rangle _{\mathrm{c}}=-\frac{\Gamma ((D+1)/2)(\eta /\alpha
)^{D+1}}{\pi ^{\frac{D+1}{2}}(a^{2}+h^{2})^{\frac{D+1}{2}}}%
\sum_{n=1}^{\infty }\frac{\cos \left( n\tilde{\alpha}_{p}\right) }{n^{D+1}}.
\label{Tiicc}
\end{equation}%
For the diagonal components of the energy-momentum tensor in the subspace $%
(x^{D-1},x^{D})$ we find%
\begin{eqnarray}
\left\langle T_{D-1}^{D-1}\right\rangle _{\mathrm{c}} &=&\left[ 1-\frac{%
(D+1)h^{2}}{a^{2}+h^{2}}\right] \langle T_{0}^{0}\rangle _{\mathrm{c}}, 
\notag \\
\left\langle T_{D}^{D}\right\rangle _{\mathrm{c}} &=&\left[ 1-\frac{%
(D+1)a^{2}}{a^{2}+h^{2}}\right] \langle T_{0}^{0}\rangle _{\mathrm{c}}.
\label{TDDcc}
\end{eqnarray}%
Finally, the expression for the off-diagonal component is reduced to%
\begin{equation}
\left\langle T_{D}^{D-1}\right\rangle _{\mathrm{c}}=\frac{(D+1)ah}{%
a^{2}+h^{2}}\langle T_{0}^{0}\rangle _{\mathrm{c}}.  \label{Toffcc}
\end{equation}%
For a conformally coupled massless field the problem on the dS bulk is
conformally related to the corresponding problem in the locally Minkowski
spacetime, with the same parameters $a$, $h$, $\alpha _{p}$, and the VEVs
are connected by the standard formulae%
\begin{equation}
\langle \varphi \varphi ^{\dagger }\rangle _{\mathrm{c}}=\frac{\langle
\varphi \varphi ^{\dagger }\rangle _{\mathrm{c}}^{\mathrm{(M)}}}{(\alpha
/\eta )^{D-1}},\;\left\langle j_{\mathrm{(p)}}^{D}\right\rangle _{\mathrm{c}%
}=\frac{\left\langle j^{D}\right\rangle _{\mathrm{c}}^{\mathrm{(M)}}}{%
(\alpha /\eta )^{D}},\;\langle T_{i}^{i}\rangle _{\mathrm{c}}=\frac{\langle
T_{i}^{i}\rangle _{\mathrm{c}}^{\mathrm{(M)}}}{(\alpha /\eta )^{D+1}}.
\label{RelM}
\end{equation}%
In the special cases $\tilde{\alpha}_{p}=0$ and $\tilde{\alpha}_{p}=\pi $
the current density vanishes and the series in the expressions for the field
squared and energy-momentum tensor are expressed in terms of the Riemann
zeta function. Depending on the values of the parameter $\tilde{\alpha}_{p}$%
, the VEVs can be either positive or negative. In particular, the
topological contribution to the energy density is negative for an untwisted
field ($\tilde{\alpha}_{p}=0$) and positive for twisted field ($\tilde{\alpha%
}_{p}=\pi $). For some intermediate value of $\tilde{\alpha}_{p}$ the VEVs
become zero. The vacuum pressure along the direction $x^{i}$, $i=0,1,\ldots
,D-2$, is given by $-\langle T_{i}^{i}\rangle _{\mathrm{c}}$ and it is equal
to the energy density with an opposite sign. This corresponds to the
equation of state of the cosmological constant type in the subspace $%
(x^{0},x^{1},\ldots ,x^{D-2})$. That is not the case for general conformal
coupling and for massive fields.

At the early stages of the cosmological expansion one has $\tau \rightarrow
-\infty $ and $\eta $ is large. In order to find the asymptotics of the VEVs
in that limit it is convenient to introduce a new integration variable $%
u=z/\eta $ in the expressions for the VEVs. The function $F_{\nu }(z)$
becomes $F_{\nu }(u\eta )$ and its argument is large. By using the
asymptotic of the modified Bessel functions for large argument it can be
shown that $F_{\nu }(z)\approx 1/z$ for $z\gg 1$. This asymptotic coincides
with the exact expression for a conformally coupled massless scalar field.
Replacing $F_{\nu }(z)=1/z$ in the expressions for the field squared,
current density, and off-diagonal component $\left\langle
T_{D}^{D-1}\right\rangle _{\mathrm{c}}$, we see that the leading terms in
the expansion over $1/\eta $ coincide with the corresponding expressions for
a conformally coupled massless field, given by (\ref{jDcc}) and (\ref{Toffcc}%
). In the expression (\ref{T00}) for the energy density, in the leading
order, one has $F^{(0)}(z)\approx -2z$ and the corresponding asymptotic,
again, coincides with the result (\ref{Tiicc}) for $i=0$. In the components (%
\ref{TDm}) and (\ref{TDD}) we have $F(z)\approx -2D\left( \xi -\xi
_{D}\right) /z$ and the terms involving the function $F_{\nu }(z)\approx 1/z$
contain additional factor $\eta ^{2}$. Hence, the latter term dominates in
the asymptotic and the leading terms coincide with (\ref{TDDcc}). We
conclude that in the limit $\tau \rightarrow -\infty $, corresponding to $%
t\rightarrow -\infty $, the leading asymptotics of the topological
contributions of the VEVs coincide with the corresponding result for a
conformally coupled massless field and the effects of gravity on those
contributions are weak. In the limit under consideration the dominant
contribution to the total VEV (\ref{TikComp}) comes from the topological
part.

The late stages of the expansion correspond to $t\rightarrow +\infty $ and $%
\eta \rightarrow 0$. Again, introducing a new integration variable $u=z/\eta 
$, we expand the function $F_{\nu }(u\eta )$ for small values of the
argument. For $\nu >0$ one has $F_{\nu }(u\eta )\propto (u\eta )^{-2\nu }$
and the topological terms in the VEVs tend to zero monotonically, like $\eta
^{D-2\nu }$ for the VEVs of the field squared and energy-momentum tensor and
like $\eta ^{D+2-2\nu }$ for the current density. For purely imaginary $\nu $%
, $\nu =i|\nu |$, and for small $\eta $ we have $F_{\nu }(u\eta )\approx 
\mathrm{Re}\,[(2/u\eta )^{2\nu }\Gamma (\nu )/\Gamma (1-\nu )]$. In this
case the topological VEVs tend to zero with oscillating behavior. The
amplitudes of the oscillations decay as $\eta ^{D}$ for the field squared
and energy-momentum tensor and as $\eta ^{D+2}$ in the case of the current
density.

\section{Conclusion}

\label{sec:Conc}

We have studied the topological Casimir effect in models with compact
dimension along which the field operator obeys helical periodicity condition
given by (\ref{hc}). A general background is considered with the metric
tensor invariant under the rotations in the plane $(x^{D-1},x^{D})$. In
addition, the presence of a gauge field is assumed with constant covariant
components $A_{D-1}$ and $A_{D}$. We can pass to the new gauge with zero
values of those components. In that gauge the field operator obeys the
helical condition (\ref{hcn}) with the new phase (\ref{alft}) depending on
the components $A_{D-1}$ and $A_{D}$. The corresponding contribution can be
interpreted in terms of the magnetic flux enclosed by the compact dimension.
We have shown that by the coordinate transformation (\ref{xbar}) the problem
with helical periodicity condition is reduced to the problem with standard
quasiperiodicity condition (\ref{qpc}) with the same phase. The length of
the corresponding compact dimension is expressed as $\sqrt{a^{2}+h^{2}}$.

The procedure we have described allows to find the VEVs of physical
observables in the topological Casimir effect for helical periodicity
conditions by using the corresponding results for quasiperiodic conditions.
That is done by the standard transformation of the tensors under the
coordinate transformation (\ref{xinv}). As important local characteristics
of the vacuum state we have considered the VEVs of the current density and
energy-momentum tensor. Their transformation laws are given by (\ref{jD}), (%
\ref{TiD}), and (\ref{TDD}). As an example of general prescription the
locally dS spacetime is considered with a single compact dimension $x^{D}$
and helicity shift along the direction $x^{D-1}$. The geometry is described
by the line element (\ref{ds2b}). The corresponding problem with general
number of toroidally compactified dimensions has been considered in \cite%
{Bell13,Bell08}. In \cite{Bell08} the VEVs of the field squared and
energy-momentum tensor were studied for periodic and antiperiodic
conditions. The VEV of the current density in the case of quasiperiodic
conditions with general phases is considered in \cite{Bell13}.

In the problem at hand the properties of the vacuum state are encoded in
two-point functions describing the correlations of the vacuum fluctuations
in different spacetime points. As a two-point function we have taken the
Hadamard function. In the problem with helical condition in locally dS
spacetime that function is expressed as (\ref{Gb}). As local characteristics
of the scalar vacuum we have considered the expectation values of the field
squared, current density and energy-momentum tensor. In the corresponding
expressions the parts induced by the compactification are explicitly
separated. The field squared and energy-momentum tensor are even functions
of the phase $\tilde{\alpha}_{p}$ in the periodicity condition, whereas the
current density is an odd function. An important difference of the helical
compactification is the presence of nonzero off-diagonal component $%
\left\langle T_{D}^{D-1}\right\rangle _{\mathrm{c}}$ of the energy-momentum
tensor. At the early stages of the dS expansion the VEVs are dominated by
the topological contribution and at those stages the influence of gravity on
the local characteristics is weak. The corresponding asymptotics are
conformally related to the VEVs on the locally Minkowski bulk. At late
stages, depending on the parameter $\nu $, the topological parts in the VEVs
decay monotonically or oscillatory and the pure dS contributions dominate.

\section*{Acknowledgments}

The work was supported by the grant No. 21AG-1C047 of the Higher Education
and Science Committee of the Ministry of Education, Science, Culture and
Sport RA.


\begin{thebibliography}{99}
\bibitem{Most97} V.M. Mostepanenko, N.N. Trunov, The Casimir Effect and Its
Applications (Clarendon, Oxford, 1997).

\bibitem{Eliz94} E. Elizalde, S.D. Odintsov, A. Romeo, A.A. Bytsenko, S.
Zerbini, Zeta Regularization Techniques with Applications (World Scientific,
Singapore, 1994).

\bibitem{Milt02} K.A. Milton, The Casimir Effect: Physical Manifestation of
Zero-Point Energy (World Scientific, Singapore, 2002).

\bibitem{Byts03} A.A. Bytsenko, G. Cognola, E. Elizalde, V. Moretti, S.
Zerbini, Analytic Aspects of Quantum Fields (World Scientific, Singapore,
2003).

\bibitem{Bord09} M. Bordag, G.L. Klimchitskaya, U. Mohideen, V.M.
Mostepanenko, Advances in the Casimir Effect (Oxford University Press,
Oxford, 2009).

\bibitem{Casi11} Lecture Notes in Physics: Casimir Physics, edited by D.
Dalvit, P. Milonni, D. Roberts, F. da Rosa (Springer, Berlin, 2011), Vol.
834.

\bibitem{Khan14} F.C. Khanna, A.P.C. Malbouisson, J.M.C. Malbouisson, A.E.
Santana, Phys. Rep. \textbf{539}, 135 (2014).

\bibitem{Duff86} M.J. Duff, B.E.W. Nilsson, C. N. Pope, Phys. Rep. \textbf{%
130}, 1 (1986).

\bibitem{Camp90} R. Camporesi, Phys. Rep. \textbf{196}, 1 (1990).

\bibitem{Byts96} A.A. Bytsenko, G. Cognola, L. Vanzo, S. Zerbini, Phys. Rep. 
\textbf{266}, 1 (1996).

\bibitem{Ford79} L.H. Ford, T. Yoshimura, Phys. Lett. A \textbf{70}, 89
(1979).

\bibitem{Sche79} J. Scherk, J.H. Schwarz, Nucl. Phys. B \textbf{153}, 61
(1979).

\bibitem{Ford80} L.H. Ford, Phys. Rev. D \textbf{22}, 3003 (1980).

\bibitem{Toms80} D.J. Toms, Phys. Rev. D \textbf{21}, 928 (1980).

\bibitem{Toms80b} D.J. Toms, Phys. Rev. D \textbf{21}, 2805 (1980).

\bibitem{Cand84} P. Candelas, S. Weinberg, Nucl. Phys. B \textbf{237}, 397
(1984).

\bibitem{Odin80} S.D. Odintsov, Sov. J. Nucl. Phys. \textbf{48}, 1148 (1988).

\bibitem{Hoso89} Y. Hosotani, Annals Phys. \textbf{190}, 233 (1989).

\bibitem{Buch89} I.L. Buchbinder, S.D. Odintsov, Int. J. Mod. Phys. A 
\textbf{04}, 4337 (1989).

\bibitem{Buch89b} I.L. Buchbinder, S.D. Odintsov, Fortschr. Phys. \textbf{37}%
, 225 (1989).

\bibitem{Quir03} M. Quiros, arXiv:hep-ph/0302189.

\bibitem{Lind04} A. Linde, J. Cosmol. Astropart. Phys. \textbf{10} (2004)
004.

\bibitem{Cao13} C.J. Cao, M. van Caspel, A.R. Zhitnitsky, Phys. Rev. D 
\textbf{87}, 105012 (2013).

\bibitem{Abre13} L.M. Abreu, C.A. Linhares, A.P.C. Malbouisson, J.M.C.
Malbouisson, Phys. Rev. D \textbf{88}, 107701 (2013).

\bibitem{Isha78} C.J. Isham, Proc. R. Soc. Lond. A \textbf{362}, 383 (1978);
Proc. R. Soc. Lond. A \textbf{364}, 591 (1978).

\bibitem{Bana79} R. Banach, J.S. Dowker, J. Phys. A \textbf{12}, 2527
(1979); J. Phys. A \textbf{12}, 2545 (1979).

\bibitem{Bell10} S. Bellucci, A.A. Saharian, V.M. Bardeghyan, Phys. Rev. D 
\textbf{82}, 065011 (2010).

\bibitem{Bell13b} S. Bellucci, A.A. Saharian, Phys. Rev. D \textbf{87},
025005 (2013).

\bibitem{Beze13} E.R. Bezerra de Mello, A.A. Saharian, Phys. Rev. D \textbf{%
87}, 045015 (2013).

\bibitem{Bell13} S. Bellucci, A.A. Saharian, H. A. Nersisyan, Phys. Rev. D 
\textbf{88}, 024028 (2013).

\bibitem{Bell15} S. Bellucci, A.A. Saharian, N.A. Saharyan, Eur. Phys. J. C 
\textbf{75}, 378 ( 2015).

\bibitem{Beze15} E.R. Bezerra de Mello, A.A. Saharian, V. Vardanyan, Phys.
Lett. B \textbf{741}, 155 (2015).

\bibitem{Bell17AdS} S. Bellucci, A.A. Saharian, V. Vardanyan, Phys. Rev. D\ 
\textbf{96}, 065025 ( 2017).

\bibitem{Bell15AdS} S. Bellucci, A.A. Saharian, V. Vardanyan, JHEP \textbf{11%
} (2015) 092.

\bibitem{Bell16AdS} S. Bellucci, A.A. Saharian, V. Vardanyan, Phys. Rev. D 
\textbf{93}, 084011 (2016).

\bibitem{Bell18AdS} S. Bellucci, A.A. Saharian, D.H. Simonyan, V.V.
Vardanyan, Phys. Rev. D\ \textbf{98}, 085020 (2018).

\bibitem{Bell20AdS} S. Bellucci, A.A. Saharian, H.G. Sargsyan, V.V.
Vardanyan, Phys. Rev. D 101, 045020 (2020).

\bibitem{Saha24} A.A. Saharian, Symmetry \textbf{16}, 92 (2024).

\bibitem{Dien01} K.R. Dienes, Phys. Rev. Lett. \textbf{88}, 011601 (2001).

\bibitem{Dien02} K.R. Dienes, A. Mafi, Phys. Rev. Lett. \textbf{88}, 111602
(2002).

\bibitem{Feng10} C.-J. Feng, X.-Z. Li, Phys. Lett. B \textbf{691}, 167
(2010).

\bibitem{Zhai11} X.-H. Zhai, X.-Z. Li, C.-J. Feng, Mod. Phys. Lett. A 
\textbf{26}, 669 (2011).

\bibitem{Oiko11} V.K. Oikonomou, Commun. Theor. Phys. \textbf{55}, 101
(2011).

\bibitem{Zhai11b} X.-H. Zhai, X.-Z. Li, C.-J. Feng, Mod. Phys. Lett. A 
\textbf{26}, 1953 (2011).

\bibitem{Zhai11c} X.-H. Zhai, X.-Z. Li, C.-J. Feng, Eur. Phys. J. C \textbf{%
71}, 1654 (2011).

\bibitem{Alei21} G. Aleixo, H.F. Santana Mota, Phys. Rev. D \textbf{104},
045012 (2021).

\bibitem{Fari22} A.J.D. Farias Junior, H.F. Santana Mota, Int. J. Mod. Phys.
D \textbf{31}, 2250126 (2022).

\bibitem{Saha23} A.A. Saharian, D.H. Simonyan, H.H. Mikayelyan, A.A.
Vantsyan, J. Contemp. Phys. \textbf{58}, 341 (2023).

\bibitem{Bell14} S. Bellucci, E. R. Bezerra de Mello, A. A. Saharian, Phys.
Rev. D \textbf{89}, 085002 (2014).

\bibitem{Saha08b} A.A. Saharian, M.R. Setare, Phys. Lett. B \textbf{659},
367 (2008).

\bibitem{Bell08} S. Bellucci, A.A. Saharian, Phys. Rev. D \textbf{77},
124010 (2008).

\bibitem{Saha09} A.A. Saharian, Int. J. Mod. Phys. A \textbf{24}, 1813
(2009).

\bibitem{Abra72} Handbook of Mathematical Functions, edited by M. Abramowitz
and I.A. Stegun (Dover, New York, 1972).

\bibitem{Prud2} A.P. Prudnikov, Y.A. Brychkov, O.I. Marichev, Integrals and
Series (Gordon and Breach, New York, NY, USA, 1986), Volume 2.
\end{thebibliography}
\end{document}